\documentclass[aps,preprint,floats,epsf,epsfig,nofootinbib,letter]{revtex4}

\usepackage{color}
\usepackage{graphicx}
\usepackage{dcolumn}
\usepackage{bm}

\def\ol{\overline}
\def\ov{\overline}
\def\be{\begin{eqnarray}}
\def\en{\end{eqnarray}}
\def\non{\nonumber}
\def\CP{{\it CP}~}
\def\vma{{_{V-A}}}

\def\la{\langle}
\def\ra{\rangle}
\def\A{{\cal A}}
\def\B{{\cal B}}

\def\PE{{P\!E}}
\def\PA{{P\!A}}

\begin{document}

\font\el=cmbx10 scaled \magstep2{\obeylines\hfill June, 2012}
\vskip 1.0 cm

\title{SU(3) symmetry breaking and {\it CP} violation in $D\to PP$ decays}

\author{Hai-Yang Cheng}
\affiliation{Institute of Physics, Academia Sinica, Taipei, Taiwan 11529, ROC}

\author{Cheng-Wei Chiang}
\affiliation{Department of Physics and Center for Mathematics and Theoretical Physics,
National Central University, Chungli, Taiwan 32001, ROC}
\affiliation{Institute of Physics, Academia Sinica, Taipei, Taiwan 11529, ROC}
\affiliation{Physics Division, National Center for Theoretical Sciences, Hsinchu, Taiwan 30013, ROC}

\bigskip
\begin{abstract}
\bigskip
Evidence of \CP violation in the charm sector has been observed recently by the LHCb and CDF Collaborations.  Adopting the topological diagram approach, we study flavor SU(3) symmetry breaking effects in the weak decay tree amplitudes of singly Cabibbo-suppressed $D\to PP$ decays.  The symmetry breaking in the color-allowed and color-suppressed amplitudes is estimated with the help of the factorization ansatz, while that in the $W$-exchange amplitude is done by fitting to related branching fraction data.  We find that the $W$-exchange amplitudes stay in the second quadrant relative to the color-allowed tree amplitude, albeit there are two possibilities for one type of $W$-exchange amplitude.  The weak decay penguin amplitudes, on the other hand, are evaluated within the framework of QCD factorization.  Using the input of topological tree amplitudes extracted from the Cabibbo-favored decay modes and the perturbative results for QCD penguin amplitudes, we make predictions for the branching fractions and \CP asymmetries of singly Cabibbo-suppressed modes.  The predictions of branching fractions are generally improved from those in the SU(3) limit.  We conclude that the direct \CP asymmetry difference between $D^0 \to K^+ K^-$ and $D^0 \to \pi^+ \pi^-$ is  about $-(0.139\pm 0.004)\%$ and $-(0.151\pm 0.004)\%$ for the two solutions of $W$-exchange amplitudes, respectively. We also find that the \CP asymmetry of $D^0\to K^0\ov K^0$ dominated by the interference between $W$-exchange amplitudes ranges from $-0.62\times 10^{-3}$ to $-1.82\times 10^{-3}$.  We study phenomenological implications of two new physics scenarios for explaining the observed \CP asymmetry in the charm sector, one with large penguin amplitudes and the other with a large chromomagnetic dipole operator.  We find that the two scenarios can be discriminated by the measurements of \CP asymmetries of a set of decay modes.
\end{abstract}



\maketitle
\small

\section{Introduction \label{sec:intro}}

Recently, the LHCb Collaboration has reported a result of a nonzero value for the difference between the time-integrated \CP asymmetries of the decays $D^0\to K^+K^-$ and $D^0\to\pi^+\pi^-$ \cite{LHCb}
\be
\Delta A_{CP}\equiv A_{CP}(K^+K^-)-A_{CP}(\pi^+\pi^-)=-(0.82\pm0.21\pm0.11)\% \qquad {\rm (LHCb)}
\en
based on 0.62 fb$^{-1}$ of 2011 data. The significance of the measured deviation from zero is 3.5$\sigma$. This first evidence of \CP violation in the charm sector was later confirmed by the CDF Collaboration based on a data sample corresponding to the integrated luminosity of $9.7$ fb$^{-1}$ \cite{CDF2}
\be
\Delta A_{CP}=-(0.62\pm0.21\pm0.10)\% \qquad {\rm (CDF)} \ .
\en
The time-integrated asymmetry can be written to first order as
\be
A_{CP}(f)=a_{CP}^{\rm dir}(f)+{\la t\ra\over\tau} a_{CP}^{\rm ind} \ ,
\en
where $a_{CP}^{\rm dir}$ is the direct \CP asymmetry, $a_{CP}^{\rm ind}$ is the indirect \CP asymmetry, $\la t\ra$ is the average decay time in the sample, and $\tau$ is the $D^0$ lifetime.
The combination of the LHCb, CDF, BaBar and Belle measurements yields
$a_{CP}^{\rm ind} = -(0.025 \pm 0.231 )\%$ and $\Delta a_{CP}^{\rm dir}=
-(0.656\pm 0.154 )\%$ \cite{HFAG}.

It is important to explore whether the first evidence of \CP violation in the charm sector is consistent with the standard model (SM) or implies new physics (NP). For some early and recent theoretical investigations, see Refs.~\cite{Quigg:1979ic,Golden:1989qx,Hinchliffe:1995hz,Bigi,BigiCDF,Ligeti,Kagan,Zhu,Rozanov,Nir,PU,Cheng:2012,Bhattacharya:2012,LiuC,Giudice,Altmannshofer,Chen,Feldmann,Li2012,Franco,Brod,Hiller,Zupan}.

It is known that a reliable theoretical description of the underlying mechanism for exclusive hadronic $D$ decays based on quantum chromodynamics (QCD) is still not yet available.  This is because the mass of the charm quark, being of order $1.5$ GeV, is not heavy enough to allow for a sensible heavy quark expansion.  Indeed, it does not make too much sense to generalize the QCD factorization (QCDF) and perturbative QCD (pQCD) approaches to charm decays as the $1/m_c$ power corrections are so large that the heavy quark expansion is beyond control.

It turns out that a more suitable framework for the analysis of hadronic charmed meson decays is the so-called topological diagram approach, a powerful tool for a model-independent analysis. In this approach, the topological diagrams are classified according to the topologies in the flavor flow of weak decay diagrams, with all strong interaction effects included implicitly.  Based on flavor SU(3) symmetry, this model-independent analysis enables us to extract the topological amplitudes by fitting to available data, to probe the relative importance of different underlying decay mechanisms, and to relate one process to another at the topological amplitude level.

The salient point of the topological diagram approach is that the magnitude and the relative strong phase of each individual topological tree amplitude in charm decays can be extracted from the data.  This allows us to calculate \CP asymmetry at tree level in a reliable way, granting us an idea about the size of \CP violation in charmed meson decays. Based on this approach, we have studied $\Delta A_{CP}$ in $D\to PP$ and $D\to VP$ decays and obtained $\Delta a_{CP}^{\rm dir}\approx -0.14\%$ and an upper bound around $-0.25\%$ in the SM \cite{Cheng:2012}. A very similar result of $\Delta a_{CP}^{\rm dir}=-0.118\%$ based on a variant of the diagrammatic approach was obtained recently in \cite{Li2012}.

While many NP models \cite{Zhu,Rozanov,Nir,LiuC,Giudice,Altmannshofer,Chen,Hiller} had been proposed to explain the measurement of large $\Delta A_{CP}$,
it was argued in \cite{Kagan,PU,Brod,Feldmann,Franco,Bhattacharya:2012} that large \CP asymmetries in singly Cabibbo-suppressed (SCS) $D$ decays were allowed in the SM and the measured $\Delta a_{CP}^{\rm dir}$ could be accommodated or marginally achieved. In particular, it was advocated by Brod {\it et al.} \cite{Brod} the possibility of a large penguin amplitude in the SM. Denote $d$- and $s$-quark penguin contractions by $P_d$ and $P_s$, respectively. Under the assumption of large enhancement in $P_{d,s}$ relative to the tree amplitude, Brod {\it et al.}
claimed that the sum of $P_d$ and $P_s$ could explain $\Delta a_{CP}^{\rm dir}$, while the difference between $P_d$ and $P_s$ explains the large disparity in the rates of $D^0\to K^+K^-$ and $D^0\to\pi^+\pi^-$. This would require that $(P_d-P_s)$ be of the same order as the tree amplitude. Anyway, the interpretation of the seemingly large SU(3) breaking in the ratio $R\equiv \Gamma(D^0\to K^+K^-)/\Gamma(D^0\to \pi^+\pi^-)$ in terms of the difference of penguin contractions $(P_d-P_s)$ is at odds with the common wisdom that the overall large SU(3) symmetry violation arises from the accumulation of several small SU(3) breaking effects in the $T$ and $E$ amplitudes.  In this work we shall carefully examine the large penguin interpretation of $R$.

As for the NP explanation of $\Delta a_{CP}^{\rm dir}$, the authors in \cite{Giudice} argued that a large chromomagnetic dipole operator could be the best NP candidate to explain the data while satisfying most flavor physics constraints at the same time.
In the present work we shall focus on the aforementioned pictures of large penguins and large chromomagnetic operator in SCS $D\to PP$ decays. In particular, we will investigate their phenomenological consequences in the \CP asymmetries of these charmed meson decay modes, seeing if there are discernible differences in the two scenarios.

Based on the topological diagram approach, we have recently studied direct \CP asymmetries in the SM for SCS $D\to PP$ and $D\to VP$ decays \cite{Cheng:2012}. In this work we will improve the previous analysis by extracting the $W$-exchange amplitudes directly from the decays $D^0\to K^+K^-, \pi^+\pi^-, \pi^0\pi^0$ and $K^0\ov K^0$ and considering explicitly all SU(3) breaking effects in SCS decay amplitudes.

The layout of the present paper is as follows.  After a brief review of the diagrammatic approach, we study various mechanisms responsible for the large SU(3) violation in the branching fraction ratio of $D^0\to K^+K^-$ to $D^0\to \pi^+\pi^-$ and fix the weak annihilation amplitudes in Section~\ref{sec:flavordiagram}. Unlike the topological tree amplitudes which can be extracted from the data, penguin amplitudes in this work are evaluated in the framework of QCD factorization as illustrated in Section~\ref{sec:penguin}. We then discuss direct \CP violation in SCS $D\to PP$ decays in Section~\ref{sec:DCPV} and possible explanations of the LHCb and CDF measurements in terms of new physics in Section~\ref{sec:NP}. Finally, Section~\ref{sec:summary} comes to our conclusions.

\section{Diagrammatic approach \label{sec:flavordiagram}}

\subsection{Topological amplitudes}

It has been established sometime ago that a least model-dependent analysis of heavy meson decays can be carried out in the so-called quark diagram (or topological diagram) approach \cite{Chau,CC86,CC87}.  In this diagrammatic scenario, the topological diagrams can be classified into three distinct groups (see Fig.~1 of \cite{Cheng:2012}). For our purposes, it suffices to consider two of them (see \cite{ChengOh} for details):

1. Tree and penguin amplitudes: color-allowed tree amplitude $T$; color-suppressed tree amplitude $C$; QCD-penguin amplitude $P$; singlet QCD-penguin amplitude $S$ involving flavor SU(3)-singlet mesons; color-favored electroweak-penguin (EW-penguin) amplitude $P_{\rm EW}$; color-suppressed EW-penguin amplitude $P_{\rm EW}^C$.

2. Weak annihilation amplitudes: $W$-exchange amplitude $E$; $W$-annihilation amplitude $A$; QCD-penguin exchange amplitude $\PE$; QCD-penguin annihilation amplitude $\PA$; EW-penguin exchange amplitude $P\!E_{\rm EW}$; EW-penguin annihilation amplitude $P\!A_{\rm EW}$.

It should be stressed that these diagrams are classified purely according to the topologies of weak interactions and flavor flows with all strong interaction effects encoded, and hence they are {\it not} Feynman graphs. In other words, all quark graphs used in this approach are topological and meant to include strong interactions to all orders, with gluon lines and quark loops included implicitly in all possible ways.  Therefore, analyses of topological graphs can provide information on final-state interactions (FSI's).

The topological amplitudes $T,C,E,A$ are extracted from the Cabibbo-favored (CF) $D\to PP$ decays to be (in units of $10^{-6}$ GeV) \cite{ChengChiang} (see also \cite{RosnerPP08})
\be \label{eq:PP1}
&& T=3.14\pm0.06, \qquad\qquad\qquad\quad
C=(2.61\pm0.08)\,e^{-i(152\pm1)^\circ}, \non \\
&&  E=(1.53^{+0.07}_{-0.08})\,e^{i(122\pm2)^\circ},
\qquad\quad  A=(0.39^{+0.13}_{-0.09})\,e^{i(31^{+20}_{-33})^\circ}
\en
for $\phi=40.4^\circ$ \cite{KLOE}, where $\phi$ is the $\eta-\eta'$ mixing angle defined in the flavor basis
\be
 \left(\matrix{ \eta \cr \eta'\cr}\right)=\left(\matrix{ \cos\phi & -\sin\phi \cr
 \sin\phi & \cos\phi\cr}\right)\left(\matrix{\eta_q \cr \eta_s
 \cr}\right),
\en
with $\eta_q={1\over\sqrt{2}}(u\bar u+d\bar d)$ and $\eta_s=s\bar s$. The fitted $\chi^2$ value is $0.29$ per degree of freedom.

The topological amplitudes $C$ and $E$ given in Eq.~(\ref{eq:PP1}) extracted from the data are much larger than those expected from naive factorization.
In the factorization approach, the color-suppressed tree amplitude $C$ has the expression
 \be \label{eq:C}
 C &=& {G_F\over
 \sqrt{2}}a_2(\ov K\pi)\,f_K(m_D^2-m_\pi^2)F_0^{D\pi}(m_K^2).
\en
Using the form factor for $D$ to $\pi$ transition determined from measurements of semileptonic $D$ meson decays to the $\pi$ meson \cite{CLEO:FF}, we find $a_2(\ov K\pi)=(0.82\pm0.02)e^{-i(152\pm1)^\circ}$ \cite{ChengChiang}, to be compared with $a_2=c_2+c_1/3\approx -0.11$ in naive factorization. Likewise, weak annihilation diagrams should be helicity suppressed, whereas data imply larger sizes for them. This is because they receive large $1/m_c$ power corrections from FSI's and large nonfactorizable contributions for $a_2$. For example, the topological amplitude $E$ receives contributions from the tree amplitude $T$ via final-state rescattering with nearby resonance effects.  The large magnitude and phase of weak annihilation can be quantitatively and qualitatively understood as elaborated on in Refs.~\cite{Zen,Chenga1a2}.

\subsection{Flavor SU(3) symmetry breaking \label{sec:su3breaking}}

Under the flavor SU(3) symmetry, one can use the topological amplitudes extracted from the CF modes to predict the rates for the SCS and doubly Cabibbo-suppressed (DCS) decays. It is known that while the agreement with experiment for the branching fractions of SCS $D\to PP$ decays is generally good (see the second column of Table~\ref{tab:BFpp}), there exist significant SU(3) breaking in some of the SCS modes in the flavor SU(3) symmetry limit.
For example, the predicted rates for the $\pi^+\pi^-$ and $\pi^0\pi^0$ modes are too large, while those for the $K^+K^-$, $\pi^+\eta^{(\prime)}$ and $K^+\eta^{(\prime)}$ modes are too small compared to data. The decay $D^0\to K^0\ov K^0$ is almost prohibited in the SU(3) symmetry limit, but the measured rate is of the same order of magnitude as that of $D^0\to \pi^0\pi^0$.

The most noticeable example of SU(3) breaking lies in the decays $D^0\to K^+K^-$ and $D^0\to \pi^+\pi^-$. Experimentally, the branching fraction of $D^0\to K^+K^-$ is larger than that of $D^0\to \pi^+\pi^-$ by a factor of $2.8$ \cite{PDG}, while their rates should be the same in the SU(3) limit. This is a long-standing puzzle since SU(3) symmetry is expected to be broken merely at the level of 30\%. Without the inclusion of SU(3) breaking effects in the topological amplitudes, we see from Table~\ref{tab:BFpp} that the predicted rate of $K^+K^-$ is even smaller than that of $\pi^+\pi^-$ due to less phase space available to the former.

The conventional wisdom for solving the above-mentioned long-standing puzzle is that the overall seemingly large SU(3) symmetry violation arises from the accumulation of several small and nominal SU(3) breaking effects in the tree amplitudes $T$ and $E$ (see e.g. \cite{Chau:SU(3)}). From the recent measurement of $\Delta A_{CP}$ by LHCb and CDF, we learn that penguin diagrams in SCS decay channels do play a crucial role for \CP violation. This leads some authors to conjecture that penguins may explain the rate disparity between $D^0\to K^+K^-$ and $D^0\to \pi^+\pi^-$.

To begin with, we write
\be \label{eq:Amp:D0pipi}
A(D^0\to \pi^+\pi^-) &=&\lambda_d(T+E+P_d+\PE_d+\PA_d)_{\pi\pi}+\lambda_s (P_s+\PE_s+\PA_s)_{\pi\pi} \non \\
&=& {1\over 2}(\lambda_d-\lambda_s)(T+E+\Delta P)_{\pi\pi}-{1\over 2}\lambda_b(T+E+\Sigma P)_{\pi\pi} \ ,
\en
where $\lambda_p\equiv V_{cp}^*V_{up}$ ($p = d,s,b$) and
\be
\Delta P &\equiv& (P_d+\PE_d+\PA_d)-(P_s+\PE_s+\PA_s)\ , \non \\
\Sigma P &\equiv& (P_d+\PE_d+\PA_d)+(P_s+\PE_s+\PA_s) \ .
\en
Likewise,
\be
A(D^0\to K^+K^-) &=&\lambda_d(P_d+\PE_d+\PA_d)_{_{K\!K}}+\lambda_s (T+E+P_s+\PE_s+\PA_s)_{_{K\!K}} \non \\
&=& {1\over 2}(\lambda_s-\lambda_d)(T+E-\Delta P)_{_{K\!K}}-{1\over 2}\lambda_b(T+E+\Sigma P)_{_{K\!K}} \ .
\en
The quantities $\Delta P$ and $\Sigma P$ are denoted by ${\cal P}^t_f$ and $2{\cal P}^p_f$, respectively, in \cite{Brod}. Recently, an interesting picture has been proposed in \cite{Brod} that $\Delta P$ dominated by the difference of $d$- and $s$-quark penguin contractions of 4-quark tree operators can explain the large rate difference between $D^0\to\pi^+\pi^-$ and $D^0\to K^+K^-$, while ${1\over 2}\Sigma P$, the average of $d$- and $s$-quark penguin contractions, can account for the large direct \CP violation $\Delta a_{CP}^{\rm dir}$. Here we would like to examine the quantity $\Delta P$.

First, assuming SU(3) symmetry for the amplitudes $T$ and $E$ for the moment,
a fit to the measured branching fractions $\B(D^0\to \pi^+\pi^-)=(1.400\pm0.026)\times 10^{-3}$ and $\B(D^0\to K^+K^-)=(3.96\pm0.08)\times 10^{-3}$ \cite{PDG} yields (in units of $10^{-6}$ GeV)
\be
\Delta P = 1.54\,e^{-i202^\circ}.
\en
Therefore, $|\Delta P/T|\sim 0.5$\,. Since the real part of $\Delta P$ is negative, it is clear that $\Delta P$ contributes constructively to $K^+K^-$ and destructively to $\pi^+\pi^-$. Next, assume that $U$-symmetry breaking in the amplitudes $T+E$ follows the pattern \cite{Brod}
\be \label{eq:UbreakinT}
(T+E)_{\pi\pi}=(T+E)(1+{1\over 2}\epsilon_{_T}), \qquad (T+E)_{_{K\!K}}=(T+E)(1-{1\over 2}\epsilon_{_T}),
\en
where $\epsilon_{_T}$ is a complex parameter with $|\epsilon_{_T}|\in (0,0.3)$. It was shown in \cite{Brod} that the relation $|\Delta P/T|\sim 0.5$ still holds roughly. Hence, the large rate disparity between $D^0\to\pi^+\pi^-$ and $D^0\to K^+K^-$ is ascribed to the difference of $d$- and $s$-quark penguin contractions. In doing so, $\Delta P$ needs to be of the same order of magnitude as $T$.

However, SU(3) breaking in reality does not necessarily follow the pattern exhibited in Eq.~(\ref{eq:UbreakinT}). In the factorization approach, various topological $T$ amplitudes  have the expressions
 \be \label{eq:T}
 T_{K\pi} &=& {G_F\over
 \sqrt{2}}a_1(\ov K\pi)\,f_\pi(m_D^2-m_K^2)F_0^{DK}(m_\pi^2), \non \\
 T_{_{KK}} &=& {G_F\over
 \sqrt{2}}a_1(K\ov K)\,f_K(m_D^2-m_K^2)F_0^{DK}(m_K^2),  \\
 T_{\pi\pi} &=& {G_F\over
 \sqrt{2}}a_1(\pi\pi)\,f_\pi(m_D^2-m_\pi^2)F_0^{D\pi}(m_\pi^2), \non
 \en
for CF $D^0\to K^-\pi^+$, SCS $D^0\to K^+K^-$ and $D^0\to\pi^+\pi^-$ decays, respectively. Hence,
\be
{T_{_{K\!K}}\over T_{\pi\pi}}={a_1(K\ov K)\over a_1(\pi\pi)}\,{f_K\over f_\pi}\,{F_0^{DK}(m_K^2)\over F_0^{D\pi}(m_\pi^2)}\,{m_D^2-m_K^2\over m_D^2-m_\pi^2}=1.32\,{a_1(K\ov K)\over a_1(\pi\pi)} \ ,
\en
where we have used the form-factor $q^2$ dependence determined experimentally
from Ref.~\cite{CLEO:FF}. The parameter $a_1$ has the general expression (see e.g. \cite{BN})
\be \label{eq:penguinWC}
a_1(P_1P_2) &=& \left(c_1+{c_2\over N_c}\right)+{c_2\over N_c}\,{C_F\alpha_s\over 4\pi}[V_1(P_2)+{4\pi^2\over N_c}H_1(P_1P_2)]+~{\cal O}(1/m_c)~{\rm corrections} \ ,
\en
where $V_1$ and $H_1$ are vertex corrections and hard spectator interactions, respectively. In the diagrammatic approach, the parameter $a_1$ can be extracted to be $|a_1(\ov K\pi)|=1.22\pm0.02$ \cite{ChengChiang} from the data of CF $D\to \ov K\pi$ decays. Recall that $c_1+{c_2\over 3}\approx 1.274$ at the scale $\mu=\ov m_c\approx 1.3$ GeV. Therefore, it is evident that nonfactorizable contributions and $1/m_c$ corrections to $a_1$ are rather small. It is thus reasonable to take the ratio $a_1(K\ov K)/a_1(\pi\pi)$ to be in the vicinity of unity.
Neglecting SU(3) breaking in the $W$-exchange amplitudes for the moment
({\it i.e.}, $E_{_{K\!K}}=E_{\pi\pi}$), we get \footnote{In the same spirit,
Bhattacharya et al. \cite{Bhattacharya:2012} found
$``P"\equiv\lambda_d(P+\PA)_d+\lambda_s(P+\PA)_s=(0.044\pm0.023)+i(0.141\pm0.036)$
after a fit to the rates of $D^0\to K^+K^-$, $\pi^+\pi^-$ and $\pi^0\pi^0$.
Then $\Delta P$ is obtained through the approximation $\Delta P\approx ``P"/\lambda_d\approx -``P"/\sin\theta_C$.}
\be
\Delta P=0.49\,e^{-i129^\circ} \ .
\en
Therefore, $|\Delta P/T| = {\cal O}(0.15)$ rather than ${\cal O}(1)$ in the presence of SU(3) violation in $T$ amplitudes suggested by the factorization approach.

Furthermore, SU(3) symmetry should be also broken in the $W$-exchange and penguin annihilation amplitudes. This can be seen from the observation of the decay $D^0\to K^0\ov K^0$ whose decay amplitude is given by
\be
A(D^0\to K^0\ov K^0)=\lambda_d(E_d+2\PA_d)+\lambda_s(E_s+2\PA_s) \ ,
\en
where $E_q$ refers to the $W$-exchange amplitude associated with $c\bar u\to q\bar q$ ($q=d,s$).
In the SU(3) limit, the decay amplitude is proportional to $\lambda_b$ and hence its rate is negligibly small, while experimentally $\B(D^0\to K^0\ov K^0)=(0.346\pm0.058)\times 10^{-3}$ \cite{PDG}. This implies sizable SU(3) symmetry violation in the $W$-exchange and QCD-penguin annihilation amplitudes.
Since the theoretical estimate of $\Delta P$ is small [see Eq.~(\ref{eq:Pover T}) below], we shall neglect $\Delta P$ and assume that the $T$ and $E$ amplitudes are responsible for the SU(3) symmetry breaking.
Noting that $E_{\pi\pi}=E_d$ and $E_{_{K\!K}}=E_s$ in our notation, we can fix the $W$-exchange amplitudes from the following four modes: $K^+K^-$, $\pi^+\pi^-$, $\pi^0\pi^0$ and $K^0\ov K^0$. \footnote{However, the $W$-exchange contribution is missing in the topological amplitude expression of the $D^0\to K^0\ov K^0$ decay in \cite{Bhattacharya:2012} as SU(3) symmetry is assumed for $W$-exchange but not for the penguin annihilation diagram $\PA$.}
Neglecting $\Delta P$ and $\lambda_b$ terms (see Table~\ref{tab:PPSCSamp} for the topological amplitudes of the above four decay modes), a fit to the data yields two possible solutions
\be \label{eq:EdEs}
{\rm (I)} &&  E_d=1.19\, e^{i15.0^\circ}E, \qquad E_s=0.58\, e^{-i14.7^\circ}E
\ , \non \\
{\rm  (II)} &&  E_d=1.19\, e^{i15.0^\circ}E, \qquad E_s=1.62\, e^{-i9.8^\circ}E
\ .
\en
The corresponding $\chi^2$ vanishes as these two solutions can be obtained exactly.
It has been noticed that a significant phase difference between $E_d$ and $E_s$ is needed in order to fit the data of $D^0\to K^0\ov K^0$ \cite{Wu:2005} and to account for the large rate difference between $D^0\to K^+K^-$ and $D^0\to\pi^+\pi^-$ \cite{ChengChiang}. When comparing the predicted branching fractions of $D^0\to\eta\eta$ and $D^0\to\eta\eta'$ (see Table \ref{tab:BFpp}), it appears that Solution I is slightly more preferable, but Solution II is equally acceptable.

We have argued in Ref.~\cite{ChengChiang} that the long-distance resonant contribution through the nearby state $f_0(1710)$ could account for SU(3)-breaking effects in the $W$-exchange topology. This has to do with the dominance of the scalar glueball content of $f_0(1710)$ and the chiral suppression effect in the ratio $\Gamma(f_0(1710)\to \pi\bar\pi)/\Gamma(f_0(1710)\to K\ov K)$.

To summarize, if SU(3) symmetry in $T$ and $E$ amplitudes holds or is broken in the manner depicted by Eq.~(\ref{eq:UbreakinT}), then a sizable  $\Delta P$ of the same order of magnitude as $T$ is needed to explain the data. If SU(3) violation due to decay constants, meson masses and form factors is taken into account in $T$ amplitudes so that $T_{_{K\!K}}/T_{\pi\pi}\approx 1.32$\,, it leads to $|\Delta P/T|\sim 0.15$\,. Finally, if the large rate difference between $D^0\to K^+K^-$ and $D^0\to\pi^+\pi^-$ is entirely accounted for by SU(3) violation in $T$ and $E$ amplitudes, $\Delta P$ will be negligibly small. Owing to the observation of $D^0\to K^0\ov K^0$ through $W$-exchange and penguin annihilation diagrams and the smallness of $\Delta P$ theoretically, we shall argue in this work that the last scenario is preferred.


\begin{table}
\caption{Topological amplitudes for singly Cabibbo-suppressed decays of charmed mesons to two pseudoscalar mesons where flavor SU(3) symmetry breaking effects are included. Summation over $p=d,~s$ is understood. For simplicity, flavor-singlet QCD penguin, flavor-singlet weak annihilation and electroweak penguin annihilation amplitudes have been neglected.
  \label{tab:PPSCSamp}}
  \footnotesize{
\bigskip
\begin{ruledtabular}
\begin{tabular}{l l l }
 & Mode & Representation
      \\
\hline
$D^0$
  & $\pi^+ \pi^-$ & $\lambda_d(0.96T+E_d)+\lambda_p(P_p+\PE_p+\PA_p)$  \\
  & $\pi^0 \pi^0$ & ${1\over \sqrt{2}}\lambda_d(-0.79C+E_d)+{1\over\sqrt{2}}\lambda_p(P_p+\PE_p+\PA_p)$ \\
  & $\pi^0 \eta $ & $-\lambda_d (E_d)\cos\phi-{1\over\sqrt{2}}\lambda_s(1.25 C)\sin\phi+\lambda_p(P_p+\PE_p)\cos\phi$ \\
  & $\pi^0 \eta' $ & $-\lambda_d (E_d)\sin\phi+{1\over\sqrt{2}}\lambda_s (1.25C)\cos\phi+\lambda_p(P_p+\PE_p)\sin\phi$ \\
  & $\eta\eta $ & ${1\over\sqrt{2}}\lambda_d(0.79 C+E_d)\cos^2\phi+\lambda_s(-{1\over 2}1.06C\sin 2\phi+\sqrt{2}\,E_s\sin^2\phi)$+${1\over\sqrt{2}}\lambda_p(P_p+\PE_p+\PA_p)\cos^2\phi$ \\
  & $\eta\eta' $ & ${1\over 2}\lambda_d(0.79 C+E_d)\sin 2\phi+\lambda_s({1\over \sqrt{2}}1.06C\cos 2\phi-E_s\sin 2\phi)$ +${1\over 2}\lambda_p(P_p+\PE_p+\PA_p)\sin 2\phi$ \\
  & $K^+ K^{-}$ & $\lambda_s(1.27T+E_s)+\lambda_p(P_p+\PE_p+\PA_p)$
     \\
  & $K^0 \ol{K}^{0}$ & $\lambda_d (E_d)+\lambda_s (E_s)+2\lambda_p (\PA_p)$  \\
\hline
$D^+$
  & $\pi^+ \pi^0$ & ${1\over\sqrt{2}}\lambda_d(0.96T+0.79C)$
     \\
  & $\pi^+ \eta $ & ${1\over\sqrt{2}}\lambda_d(0.82T+0.93C+1.15A)\cos\phi-\lambda_s(1.29C)\sin\phi+
  \sqrt{2}\lambda_p(P_p+\PE_p)\cos\phi$
      \\
  & $\pi^+ \eta' $ & ${1\over\sqrt{2}}\lambda_d(0.82T+0.93C+1.56A)\sin\phi+\lambda_s(1.29C)\cos\phi+
  \sqrt{2}\lambda_p(P_p+\PE_p)\sin\phi$
     \\
  & $K^+ \ol{K}^{0}$ & $\lambda_d (0.86A)+\lambda_s(1.27T)+\lambda_p(P_p+\PE_p)$
     \\
\hline
$D_s^+$
  & $\pi^+ K^{0}$ & $\lambda_d(1.12T)+\lambda_s(A)+\lambda_p(P_p+\PE_p)$
     \\
  & $\pi^0 K^{+}$ & ${1\over\sqrt{2}}[-\lambda_d(0.91C)+\lambda_s(A)+\lambda_p(P_p+\PE_p)]$
     \\
  & $K^{+}\eta$ & $\frac{1}{\sqrt{2}}\lambda_p[0.94C\delta_{pd} + A\delta_{ps}
  + P_p+P\!E_p ]\cos\phi-\lambda_p[(1.28T+1.24C+A)\delta_{ps} + P_p+P\!E_p]\sin\phi$
  \\
  & $K^{+}\eta' $ & $\frac{1}{\sqrt{2}}\lambda_p[0.94C\delta_{pd} + A\delta_{ps}
  + P_p+P\!E_p]\sin\phi+\lambda_p[(1.28T+1.24C+A)\delta_{ps} + P_p+P\!E_p] \cos\phi$
   \\
\end{tabular}
\end{ruledtabular} }
\end{table}
%



\begin{table}
\caption{Branching fractions (in units of $10^{-3}$) of singly Cabibbo-suppressed $D\to PP$ decays.
The column denoted by $\B_{_{\rm SU(3)}}$ shows the predictions based on our best-fitted results in Eq.~(\ref{eq:PP1}) with exact flavor SU(3) symmetry, while SU(3) symmetry breaking effects are taken into account in the column denoted by $\B_{_{\rm SU(3)\!-\!breaking}}$. The first (second) entry in $D^0\to \eta\eta$, $\eta\eta'$, $K^+K^-$ and $K^0\ov K^0$ modes is for Solution I (II) of $E_d$ and $E_s$ in Eq.~(\ref{eq:EdEs}). Experimental results of branching fractions are taken from Ref.~\cite{PDG}.
  \label{tab:BFpp}}
\bigskip
\begin{tabular}{l c c c} \hline\hline
Decay Mode & $\B_{_{\rm SU(3)}}$ & $\B_{_{\rm SU(3)\!-\!breaking}}$ & $\B_{\rm expt}$ \\
 \hline
$D^0\to \pi^+ \pi^-$~~~~ & ~~$2.26\pm0.13$~~ & ~~$1.40\pm0.11$ & ~~$1.400\pm0.026$ \\
$D^0\to\pi^0 \pi^0$  & $1.35\pm0.09$ & ~~$0.78\pm0.06$ & ~~$0.80\pm0.05$  \\
$D^0\to \pi^0 \eta $  & $0.75\pm0.05$ & ~~$0.83\pm0.06$ & ~~$0.68\pm0.07$  \\
$D^0\to \pi^0 \eta' $  & $0.75\pm0.05$ & ~~$1.42\pm0.08$ & ~~$0.89\pm0.14$  \\
$D^0\to \eta\eta $ & $1.43\pm0.09$ & ~~$1.68\pm0.09$ & ~~$1.67\pm0.20$ \\
                   & $1.43\pm0.09$ & ~~$1.89\pm0.10$ & \\
$D^0\to \eta\eta' $  & $1.20\pm0.10$ & ~~$0.68\pm0.06$ & ~~$1.05\pm0.26$  \\
                     & $1.20\pm0.10$ & ~~$2.11\pm0.20$ & \\
$D^0\to K^+ K^{-}$   & $1.89\pm0.11$ & ~~$3.89\pm0.16$ & ~~$3.96\pm0.08$  \\
                     & $1.89\pm0.11$ & ~~$3.90\pm0.22$ & \\
$D^0\to K^0 \ol{K}^{0}$  & 0 & ~~$0.346\pm0.034$ & ~~$0.346\pm0.058$ \\
                         & 0 & ~~$0.345\pm0.034$ & \\
$D^+\to \pi^+ \pi^0$   & $0.88\pm0.06$ & ~~$0.97\pm0.07$ & ~~$1.19\pm0.06$ \\
$D^+\to \pi^+ \eta $  & $1.59\pm0.35$ & ~~$3.35\pm0.39$ & ~~$3.53\pm0.21$   \\
$D^+\to\pi^+ \eta' $  & $3.68\pm0.33$ & ~~$4.62\pm0.31$ & ~~$4.67\pm0.29$   \\
$D^+\to K^+ \ol{K}^{0}$  & $5.46\pm0.55$ & ~~$8.93\pm0.85$ & ~~$5.66\pm0.32$ \\
$D_s^+\to \pi^+ K^{0}$  & $2.85\pm0.28$ & ~~$3.65\pm0.33$ & ~~$2.42\pm0.16$\\
$D_s^+\to \pi^0 K^{+}$  & $0.73\pm0.09$ & ~~$0.73\pm0.09$ & ~~$0.62\pm0.21$  \\
$D_s^+\to K^{+}\eta$   & $0.79\pm0.08$ & ~~$0.84\pm0.08$ & ~~$1.75\pm0.35$  \\
$D_s^+\to K^{+}\eta' $  & $1.02\pm0.17$ & ~~$1.25\pm0.20$ & ~~$1.8\pm0.6$  \\
\hline\hline
\end{tabular}
\end{table}
%


To estimate the effects of SU(3) symmetry violation in $T$ and $C$ amplitudes, we shall rely on the factorization ansatz. In this approach, the topological amplitudes $T$ and $C$ extracted from
CF $D\to \bar K\pi$ decays have the expressions
 \be \label{eq:T}
 T &=& {G_F\over
 \sqrt{2}}a_1(\ov K\pi)\,f_\pi(m_D^2-m_K^2)F_0^{DK}(m_\pi^2),
 \en
and Eq. (\ref{eq:C}) for the amplitude $C$.
In \cite{ChengChiang} we have illustrated SU(3) breaking effects in some selective SCS modes. For example, the relevant factorizable amplitudes for $D^+\to\pi^+\eta^{(')}$ decays are
\be
T_{\pi\eta_q} &=& {G_F\over\sqrt{2}}\,a_1 f_\pi(m_D^2-m_{\eta_q}^2)F_0^{D\eta_q}(m_\pi^2), \non \\
C_{\pi\eta_q} &=& {G_F\over\sqrt{2}}\,a_2 f_{q}(m_D^2-m_\pi^2)F_0^{D\pi}(m_{\eta_q}^2), \non \\
C_{\pi\eta_s} &=& {G_F\over\sqrt{2}}\,a_2 f_{s}(m_D^2-m_\pi^2)F_0^{D\pi}(m_{\eta_s}^2),
\en
where $f_q$, $f_s$ are the decay constants of $\eta_q$ and $\eta_s$, respectively. We shall use the parameters $f_q=f_\pi$, $f_s=1.325f_\pi$, $m_{\eta_q}=741$ MeV and $m_{\eta_s}=783$ MeV \cite{FKS}, and assume the form factor $F_0^{D\eta_q}$ to be the same as $F_0^{D\pi}$. As pointed out in \cite{ChengChiang}, one needs SU(3) violation in weak annihilation to get a better agreement with experiment. For this purpose, we rely on the decay constants involved in the processes to estimate the SU(3) breaking effects in the $W$-annihilation amplitude $A$. In CF $D\to PP$ decays, the topological amplitude $A$ is extracted from $D_s^+\to K^+\ov K^0$ which involves the decay constants $f_{D_s}$ and $f_K$.
For $D^+\to\pi^+\eta$, we thus have $A_{\pi\eta}=(f_{D}/f_{D_s})(f_\pi f_q/f_K^2)A=1.15A$ and, likewise, $A_{\pi\eta'}=(f_{D}/f_{D_s})(f_\pi f_s/f_K^2)A=1.56A$, where use of world averages $f_D=213$ MeV and $f_{D_s}=248$ MeV \cite{Laiho} has been made. Finally, the decay amplitudes read (see also Table \ref{tab:PPSCSamp})
\be \label{eq:Ddpieta}
\A(D^+\to\pi^+\eta) &=&{1\over\sqrt{2}}V_{cd}^*V_{ud}(0.816\,T+0.930\,C+1.15A)\cos\phi-V_{cs}^*V_{us}1.285\, C\sin\phi, \non \\
\A(D^+\to\pi^+\eta') &=&{1\over\sqrt{2}}V_{cd}^*V_{ud}(0.816\,T+0.930\,C+1.56A)\sin\phi+V_{cs}^*V_{us}1.285\, C\cos\phi.
\en
From Table \ref{tab:BFpp} we see that the discrepancy between theory and experiment for $\B(D^+\to\pi^+\eta^{(')})$ is greatly improved.

SU(3) breaking effects in the topological amplitudes for SCS $D\to PP$ decays are summarized in Table~\ref{tab:PPSCSamp}. Theory predicted and measured branching fractions are given in Table~\ref{tab:BFpp}. \footnote{Our previous result of $a_{\rm dir}^{\rm (tot)}(D_s^+\to K^+\eta)$  is erroneous and it has been corrected in the erratum of \cite{Cheng:2012}.}
While the agreement with experiment is improved for most of the SCS modes after taking into account SU(3) breaking effects in decay amplitudes, there are a few exceptions. For example, the predicted rate for $D^+\to K^+\ov K^0$ becomes even worse compared to the prediction based on SU(3) symmetry. It is possible that the effective parameter $a_1(D^+\to K^+\ov K^0)$ is smaller than $a_1(D^0\to \ov K\pi)$.  Finally, we note in passing that the central values of the predicted branching fraction in the SU(3) limit given in Table~\ref{tab:BFpp} are sometimes slightly different from those given in Table~II of Ref.~\cite{ChengChiang}.  This is  because the current work adds penguin contributions into the analysis.  The errors associated with these branching fractions are larger than before because a Monte Carlo sampling method is used here for error estimation, instead of the simple error propagation method used in Ref.~\cite{ChengChiang}.  The same Monte Carlo method is used throughout this paper for error estimation.

\section{Penguin amplitudes in QCD factorization \label{sec:penguin}}

Although the topological tree amplitudes $T,C,E$ and $A$ for hadronic $D$ decays can be extracted from the data, we still need information on penguin amplitudes (QCD penguin, penguin exchange and penguin annihilation) in order to estimate \CP violation in the SCS decays. Unlike the tree amplitudes, it is more difficult to extract the topological penguin amplitudes reliably from the data.  This is because the use of the topological approach relies heavily on SU(3) symmetry which leads to negligible penguin amplitudes in $D$ decays.  Consequently, the extraction of penguin amplitudes depends on SU(3) breaking effects in tree amplitudes.  Indeed, we have shown in Sec.~\ref{sec:su3breaking} that the difference in penguin contractions $\Delta P$ is sensitive to how SU(3) symmetry breaking is treated in the tree amplitudes $T$ and $E$.  Therefore, we shall rely on theory to estimate the penguin contribution.

With the advent of heavy quark effective theory, it is known that nonleptonic $B$ decays can be analyzed systematically within the QCD framework.  There are three popular approaches available in this regard: QCD factorization (QCDF) \cite{BBNS99}, perturbative QCD (pQCD) \cite{pQCD} and soft-collinear effective theory (SCET) \cite{SCET}. QCDF is phenomenologically quite successful in describing charmless hadronic $B$ decays (see Ref.~\cite{Cheng2009} for the QCDF predictions of $B_{u,d,s}\to PP,VP$ and $VV$ decays and the comparison with experiment and the theory predictions of pQCD and SCET).  This indicates that the dynamics of the penguin mechanism in penguin-dominated $B$ decays is well described by QCDF.

As noted in Sec.~\ref{sec:intro}, since the charmed quark is not heavy enough and $1/m_c$ power corrections are so large, a sensible heavy quark expansion in $1/m_c$ is no longer applicable to aforementioned QCD-inspired approaches.  This means that the less sophisticated approaches such as the naive factorization or the improved version of factorization such as the generalized factorization \cite{Ali,Cheng99} can serve as a reasonable starting point. However, none of the existing theoretical frameworks work well for hadronic $D$ decays as the large $1/m_c$ power corrections are nonperturbative in nature and cannot be tackled using the factorization approach.  Nevertheless, in this work we shall apply QCDF to the zeroth order in the power expansion of $\Lambda_{\rm QCD}/m_c$ (except for chirally enhanced power corrections characterized by the chiral factor $r_\chi$ given in Eq.~(\ref{eq:rchi}) which are formally $1/m_c$ suppressed but numerically very important) to estimate penguin amplitudes in charm decays for the following reasons: (i) the expressions of penguin contractions in different approaches are similar except that the penguin loop function $G(s,x)$ to be defined below is convoluted with the light-cone distribution amplitude of the light meson in QCDF, while it is replaced by $G(s,k^2)$ where $k^2$ is the squared momentum carried by the virtual gluon; and (ii) vertex corrections can provide a strong phase which is absent in the generalized factorization approach.  Hence, we shall work in QCDF \cite{BBNS99,BN} to evaluate perturbative penguin amplitudes, but keep in mind that we employ this approach simply for a crude estimate of the penguin contractions.  As for power corrections to QCD-penguin exchange amplitude $\PE$ and QCD-penguin annihilation amplitude $\PA$, we shall adopt a different strategy to deal with long-distance effects due to FSI's, as will be elaborated in Sec.~\ref{sec:CPV-from-penguin}.

To calculate the penguin contributions, we start from the short-distance effective Hamiltonian
\be
{\cal H}_{\rm eff}={G_F\over\sqrt{2}}\left[\sum_{p=d,s}\lambda_p(c_1O_1^p+c_2O_2^p+c_{8g}O_{8g})
-\lambda_b\sum_{i=3}^{6}c_iO_i\right] \ ,
\en
where
\be
&& O_1^p=(\bar pc)_{_{V-A}}(\bar up)_\vma, \qquad\qquad\quad O_2^p=(\bar p_\alpha c_\beta)_{_{V-A}}(\bar u_\beta p_\alpha)_\vma, \non \\
&& O_{3(5)}=(\bar uc)\vma\sum_q(\bar qq)_{_{V\mp A}}, \qquad~~
O_{4(6)}=(\bar u_\alpha c_\beta)_\vma\sum_q (\bar q_\beta q_\alpha)_{_{V\mp A}},  \non \\
&& O_{8g}=-{g_s\over 8\pi^2}m_c\,\bar u\sigma_{\mu\nu}(1+\gamma_5)
G^{\mu\nu}c \ ,
\en
with $O_3$--$O_6$ being the QCD penguin operators and $(\bar q_1q_2)_{_{V\pm A}}\equiv\bar q_1\gamma_\mu(1\pm \gamma_5)q_2$. The electroweak penguin operators are not included in the Hamiltonian as they can be neglected in practice. For the Wilson coefficients, we follow \cite{Li2012} to take $c_1=1.22$, $c_2=-0.43$, $c_3=0.018$, $c_4=-0.046$, $c_5=0.013$, $c_6=-0.044$ and $c_{8g}=-0.11$ evaluated at the scale $\mu=m_c$.

Let us first consider the penguin amplitudes in $D\to P_1P_2$ decays
\be \label{eq:PinPP}
S_{P_1P_2}&=&{G_F \over \sqrt{2}} [a_3(P_1P_2)- a_5(P_1P_2)]X^{(DP_1, P_2)} \ , \non \\
P^p_{P_1P_2} &=& {G_F \over \sqrt{2}} [a^p_4(P_1P_2)+r_\chi^{P_2} a^p_6(P_1P_2)]X^{(DP_1,P_2)} \ , \non \\
 P\!E_{P_1P_2}^p
 &=& {G_F \over \sqrt{2}}\,
   (i f_D f_{P_1} f_{P_2})\left[ b_3^p \right]_{P_1 P_2 } ~,  \\
 P\!A_{P_1P_2}^p
 &=& {G_F \over \sqrt{2}} \,
   (i f_D f_{P_1} f_{P_2})\left[ b_4^p \right]_{P_1 P_2 } ~, \non
\en
where $C_F\equiv (N_c^2-1)/(2N_c)$ with $N_c=3$,
\begin{eqnarray} \label{eq:rchi}
 r_\chi^P(\mu) = {2m_P^2 \over m_c(\mu)(m_2+m_1)(\mu)}
\end{eqnarray}
is a chiral factor and
\be \label{eq:X}
X^{(D P_1, P_2)} &\equiv& \langle P_2| J^{\mu} |0 \rangle
  \langle P_1| J'_{\mu} |D \rangle
  =i f_{P_2} (m_{D}^2 -m^2_{P_1}) ~F_0^{D P_1} (m_{P_2}^2) ~,
\en
with $p_c$ being the center-of-mass momentum of either final state particle. Here we have followed the conventional Bauer-Stech-Wirbel definition for the form factor $F_{0}^{DP}$ \cite{BSW}.
The flavor operators $a_i^p$ are basically the Wilson coefficients in conjunction with short-distance nonfactorizable corrections such as vertex corrections $V_i$, penguin contractions ${\cal P}_i$ and hard spectator interactions $H_i$:
\be \label{eq:penguinWC}
a_3(P_1P_2) &=& \left(c_3+{c_4\over N_c}\right)+{c_4\over N_c}\,{C_F\alpha_s\over 4\pi}[V_3(P_2)+{4\pi^2\over N_c}H_3(P_1P_2)],  \non \\
a^p_4(P_1P_2) &=& \left(c_4+{c_3\over N_c}\right)+{c_3\over N_c}\,{C_F\alpha_s\over 4\pi}[V_4(P_2)+{4\pi^2\over N_c}H_4(P_1P_2)]+{\cal P}^p_4(P_2),  \non \\
a_5(P_1P_2) &=& \left(c_5+{c_6\over N_c}\right)+{c_6\over N_c}\,{C_F\alpha_s\over 4\pi}[V_5(P_2)+{4\pi^2\over N_c}H_5(P_1P_2)], \non \\
a^p_6(P_1P_2) &=& \left(c_6+{c_5\over N_c}\right)+{c_5\over N_c}\,{C_F\alpha_s\over 4\pi}[V_6(P_2)+{4\pi^2\over N_c}H_6(P_1P_2)]+{\cal P}^p_6(P_2),
\en
where the explicit expressions of $V_i$ and $H_i$ can be found in \cite{BN}.
The strong phase of the QCD penguin amplitude arises from vertex corrections and penguin contractions. The order $\alpha_s$ corrections from penguin contraction read \cite{BN}
\be \label{eq:P46}
{\cal P}^p_4&=&{C_F\alpha_s\over 4\pi N_c}\Bigg\{ c_1\left[ {4\over 3}{\rm ln}{m_c\over \mu}+{2\over 3}-G_{M_2}(s_p)\right]+c_3\left[ {8\over 3}{\rm ln}{m_c\over \mu}+{4\over 3}-G_{M_2}(s_u)-G_{M_2}(1)\right]  \non \\
&& +(c_4+c_6)\left[{16\over 3}{\rm ln}{m_c\over \mu}-G_{M_2}(s_u)-G_{M_2}(s_d)-G_{M_2}(s_s)-G_{M_2}(1)\right]-2c_{8g}^{\rm eff}\int^1_0 {dx\over 1-x}\Phi_{M_2}(x) \Bigg\} \ ,  \non \\
{\cal P}^p_6&=&{C_F\alpha_s\over 4\pi N_c}\Bigg\{ c_1\left[ {4\over 3}{\rm ln}{m_c\over \mu}+{2\over 3}-\hat G_{M_2}(s_p)\right]+c_3\left[ {8\over 3}{\rm ln}{m_c\over \mu}+{4\over 3}-\hat G_{M_2}(s_u)-\hat G_{M_2}(1)\right]   \\
&& +(c_4+c_6)\left[{16\over 3}{\rm ln}{m_c\over \mu}-\hat G_{M_2}(s_u)-\hat G_{M_2}(s_d)-\hat G_{M_2}(s_s)-\hat G_{M_2}(1)\right] -2c_{8g}^{\rm eff} \Bigg\} \ ,  \non
\en
where $c_{8g}^{\rm eff}=c_{8g}+c_5$,
$s_i=m_i^2/m_c^2$, and
\be \label{eq:G}
G_{M_2}(s)=\int_0^1 dx\,G(s,1-x)\Phi_{M_2}(x), \qquad
\hat G_{M_2}(s)=\int_0^1 dx\,G(s,1-x)\Phi_{m_2}(x),
\en
with
\be
G(s,x)=-4\int^1_0 du\,u(1-u){\rm ln}[s-u(1-u)x].
\en
Here $\Phi_{M_2}$ ($\Phi_{m_2}$) is the twist-2 (-3) light-cone distribution amplitude for the meson $M_2$. In the generalized factorization approach, the perturbative loop functions $G_{M_2}(s)$ and $\hat G_{M_2}(s)$ are replaced by \be
G(s,k^2)=-4\int^1_0 du\,u(1-u){\rm ln}\left[s-u(1-u){k^2\over m_c^2}\right],
\en
where $k^2$ is the virtual gluon's momentum squared.

The annihilation operators $b_{3,4}^p$ in Eq.~(\ref{eq:PinPP}) are given by
\be \label{eq:bi}
b_3^p &=& {C_F\over N_c^2}\left[c_3A_1^i+c_5(A_3^i+A_3^f)+N_cc_6 A_3^f\right], \non \\
b_4^p &=& {C_F\over N_c^2}\left[c_4A_1^i+c_6A_2^i\right],
\en
where the annihilation amplitudes $A_{1,2,3}^{i,f}$ are defined in Ref.~\cite{BN}. As stressed in \cite{Kagan}, contributions to penguin weak annihilation amplitudes $\PE$ and $\PA$ from penguin contractions should be included in order to ensure renormalization scheme and scale independence. This amounts to the effective penguin Wilson coefficients \cite{Ali,Cheng99}
\be
c_{4,6}^{p,{\rm eff}} = c_{4,6}(\mu)+{N_c\over 2C_F}{\cal P}^p_{4,6}(\mu), \qquad
c_{3,5}^{p,{\rm eff}} = c_{3,5}(\mu)-{1\over 2C_F}{\cal P}^p_{4,6}(\mu).
\en
Hence, the penguin Wilson coefficients in Eq. (\ref{eq:bi}) should be replaced by the effective ones.

In practical calculations of QCDF, the superscript `$p$' can be omitted for $a_3$ and $a_5$. For $a_{4,6}^p$ and $c_{3-6}^{p,{\rm eff}}$, their  `$p$' dependence is dictated by the terms $G_{M_2}(s_p)$ and $\hat G_{M_2}(s_p)$ defined in Eq.~(\ref{eq:G}). Also as explained in \cite{Cheng:2012}, we shall not consider the spectator contributions to $a_i$ because the relevant scale $\mu_h\approx 0.8$ GeV is beyond the regime where perturbative QCD is applicable.

\section{Direct \CP violation \label{sec:DCPV}}

\subsection{Tree-level \CP violation}

Direct \CP asymmetry in hadronic charm decays defined by
\be
a_{CP}^{\rm dir}(f)={\Gamma(D\to f)-\Gamma(\overline D\to \bar f)\over \Gamma(D\to f)+\Gamma(\overline D\to \bar f)}
\en
can occur even at the tree level \cite{Cheng1984}. As emphasized in \cite{Cheng:2012},
the great merit of the topological approach is that the magnitude and the relative strong phase of each individual topological tree amplitude in charm decays can be extracted from the data. Hence, the estimate of $a_{dir}^{\rm (tree)}$ should be trustworthy. Larger \CP asymmetries can be achieved in those decay modes with interference between $T$ and $C$ or $C$ and $E$. For example, $a_{dir}^{({\rm tree})}$ is of order $(0.7-0.8)\times 10^{-3}$ for $D^0\to \pi^0\eta$ and $D_s^+\to K^+\eta$ (see Table~\ref{tab:CPVpp}). Direct \CP violation in $D^0\to K^0\ov K^0$ is given by
\be \label{eq:acpKK}
a_{dir}^{({\rm tree})}(D^0\to K^0\ov K^0)= {2{\rm Im}(\lambda_d\lambda_s^*)\over |\lambda_d|^2}\,
{{\rm Im}(E_d^*E_s)\over |E_d-E_s|^2}= 1.2\times 10^{-3} {|E_dE_s|\over |E_d-E_s|^2}\sin\delta_{ds} \ ,
\en
where $\delta_{ds}$ is the strong phase of $E_s$ relative to $E_d$. Substituting the two solutions for $E_d$ and $E_s$ given in Eq.~(\ref{eq:EdEs}) in the above equation yields
\be
a_{dir}^{({\rm tree})}(D^0\to K^0\ov K^0)=\left\{
\begin{array}{cl}
    -0.7\times 10^{-3}
      & \quad \mbox{Solution~I} \ , \\
    -1.7\times 10^{-3}
      & \quad \mbox{Solution~II} \ .
    \end{array}\right.
\en
For comparison, $a_{dir}^{({\rm tree})}(D^0\to K^0\ov K^0)=0.69\times 10^{-3}$ is predicted in \cite{Li2012}.

From Table~\ref{tab:CPVpp} we see that almost all the predicted tree-level \CP asymmetries in \cite{Li2012} are of opposite signs to ours. This can be traced back to the phase of the $W$-exchange amplitude. For CF $D\to PP$ decays, its phase is $(122\pm 2)^\circ$ with $\chi^2=0.29$ per degree of freedom [Eq.~(\ref{eq:PP1})]. For SCS decays, the phases of $E_d$ and $E_s$ [see Eq.~(\ref{eq:EdEs})] lie in the range of $107^\circ\sim 137^\circ$. Therefore, the $W$-exchange amplitude in this work is always in the second quadrant, while the $E$ amplitude in \cite{Li2012} lies in the third quadrant because a global fit to all the data of 28 CF and SCS $D\to PP$ branching fractions has been performed there with $\chi^2=7.3$ per degree of freedom.  As a result, the imaginary part of $E$ in \cite{Li2012} has a sign opposite to ours, and this explains the sign difference between this work and \cite{Li2012} for $a_{dir}^{({\rm tree})}$.


\begin{table}
\caption{Direct \CP asymmetries (in units of $10^{-3}$) of $D\to PP$ decays, where $a_{dir}^{({\rm tree})}$ denotes \CP asymmetry arising from purely tree amplitudes and $a_{dir}^{({\rm tot})}$ from the total amplitude. The first (second) entry in $D^0\to \eta\eta$, $\eta\eta'$, $K^+K^-$ and $K^0\ov K^0$ is for Solution I (II) of $E_d$ and $E_s$ [Eq.~(\ref{eq:EdEs})].
For QCD-penguin exchange $\PE$, we assume that it is similar to the topological $E$ amplitude [see Eq.~(\ref{eq:PE})].  World averages of experimental measurements are taken from Ref.~\cite{HFAG}. For comparison, the predicted results of $a_{dir}^{({\rm tree})}$ and $a_{dir}^{({\rm tot})}$ in \cite{Li2012} are also presented.
  \label{tab:CPVpp}}
\begin{tabular}{l c r c r c} \hline\hline
Decay Mode &  $a_{dir}^{({\rm tree})}$(this work)~ &$a_{dir}^{({\rm tree})}$\cite{Li2012}~~~
     & $a_{dir}^{({\rm tot})}$(this work)~ & $a_{dir}^{({\rm tot})}$\cite{Li2012}~~ & Expt. \\
 \hline
$D^0\to \pi^+ \pi^-$ & $0$ & 0~~~
     & $0.95\pm0.04$ & 0.68~~  & $2.0\pm2.2$ \\
$D^0\to\pi^0 \pi^0$  & $0$ & 0~~~
     & $0.80\pm0.04$ & 0.20~~ & $1\pm48$ \\
$D^0\to \pi^0 \eta $  &  $0.82\pm0.03$ & $-0.33$~~~     & $0.08\pm0.04$ & $-0.55$~~  \\
$D^0\to \pi^0 \eta' $  &  $-0.39\pm0.02$~~ & $0.54$~~~     & $0.01\pm0.02$ & 1.99~~  \\
$D^0\to \eta\eta $ &  $-0.28\pm0.01$~~ & 0.28~~~ & $-0.58\pm0.02$~~ & 0.08~~ \\
& $-0.42\pm0.02$~~ & 0.28~~~ & $-0.74\pm0.02$~~ & 0.08~~  \\
$D^0\to \eta\eta' $  &  $0.49\pm0.02$ & $-0.30$~~~     & $0.54\pm0.02$ & $-0.98$~~   \\
 &  $0.38\pm0.02$ & $-0.30$~~~     & $0.34\pm0.02$ & $-0.98$~~   \\
$D^0\to K^+ K^{-}$   &  $0$ & 0~~~
     & $-0.42\pm0.01$~~ & $-0.50$~~ & $-2.3\pm1.7$  \\
     &  $0$ & 0~~~
     & $-0.53\pm0.02$~~ & $-0.50$~~  \\
$D^0\to K^0 \ol{K}^{0}$  & $-0.73$ & 1.11~~~ & $-0.63\pm0.01$~~ & $1.37$~~  \\
& $-1.73$ & 1.11~~~ & $-1.81\pm0.01$~~ & $1.37$~~  \\
$D^+\to \pi^+ \pi^0$   & $0$ & 0~~~
     & $0$ & 0~~ & $29\pm29$  \\
$D^+\to \pi^+ \eta $  & $0.35\pm0.06$ & $-0.54$~~~ &  $-0.74\pm0.06$~~ & $-0.52$~~  &$17.4\pm11.5$ \footnotemark[1] \\
$D^+\to\pi^+ \eta' $  &  $-0.21\pm0.04$~~ & 0.39~~~    & $0.33\pm0.07$ & 1.52~~  & $-1.2\pm11.3$  \footnotemark[1] \\
$D^+\to K^+ \ol{K}^{0}$~~~  &  $-0.07\pm0.06$~~ & $-0.14$~~~   & $-0.39\pm0.04$~~ & $-1.00$~~  & $-1.0\pm5.9$ \\
$D_s^+\to \pi^+ K^{0}$  &  $0.07\pm0.06$ & $0.14$~~~     & $0.45\pm0.03$ &  $1.00$~~ & $66\pm24$ \\
$D_s^+\to \pi^0 K^{+}$  & $0.01 \pm 0.11$ & 0.33~~~
     & $0.94 \pm 0.10$ & 0.72~~ & $266\pm228$    \\
$D_s^+\to K^{+}\eta$   &  $-0.71\pm0.05$~~ & $-0.19$~~~    & $-0.61\pm0.05$~~ & 0.83~~ & $93\pm152$   \\
$D_s^+\to K^{+}\eta' $  &  $0.36\pm0.04$ & $-0.41$~~~     & $-0.28\pm0.12$~~ & $-1.78$~~  & $60\pm189$   \\
\hline\hline
\end{tabular}
\footnotetext[1]{Data from \cite{Belle}.}
\end{table}
%


\subsection{Penguin-induced \CP violation \label{sec:CPV-from-penguin}}

Direct \CP violation does not occur at the tree level in some of the SCS decays, such as $D^0\to K^+K^-$ and $D^0\to\pi^+\pi^-$. In these two decays, \CP asymmetry can only arise from the interference between tree and penguin amplitudes denoted by $a_{dir}^{\rm (t+p)}$. From Eq.~(\ref{eq:Amp:D0pipi}) we obtain
\be \label{eq:pipiacp}
a_{dir}^{({\rm t+p})}(\pi^+\pi^-) &=& {4{\rm Im}[(\lambda_d-\lambda_s)\lambda_b^*]\over |\lambda_d-\lambda_s|^2}\,
{{\rm Im}[(T^*+E^*+\Delta P^*)(T+E+\Delta P+\Sigma P-\Delta P)]_{\pi\pi}\over |T+E+\Delta P|_{\pi\pi}^2 }   \non \\
&\approx& 1.2\times 10^{-3} \left| {P_s+P\!E_s+P\!A_s\over T+E+\Delta P}\right|_{\pi\pi}\sin\delta_{\pi\pi} \ ,
\en
where $\delta_{\pi\pi}$ is the strong phase of $(P_s+P\!E_s+P\!A_s)_{\pi\pi}$ relative to $(T+E+\Delta P)_{\pi\pi}$. Likewise,
\be  \label{eq:KKacp}
a_{dir}^{({\rm t+p})}(K^+K^-)
&\approx& -1.2\times 10^{-3} \left| {P_d+P\!E_d+P\!A_d \over T+E-\Delta P}\right|_{_{K\!K}}\sin\delta_{_{K\!K}} \ ,
\en
with $\delta_{_{K\!K}}$ being the strong phase of $(P_d+P\!E_d+P\!A_d)_{_{K\!K}}$ relative to $(T+E-\Delta P)_{_{K\!K}}$. Therefore, we have the relation
\be
a_{dir}^{({\rm t+p})}(K^+K^-)=-a_{dir}^{({\rm t+p})}(\pi^+\pi^-)\quad {\rm in~SU(3)~limit.}
\en
Note that the expression of $a_{dir}^{({\rm t+p})}(\pi^+\pi^-)$  given in \cite{Cheng:2012}
\be \label{eq:pipiacp2}
a_{dir}^{({\rm t+p})}(\pi^+\pi^-) &=& {2{\rm Im}(\lambda_d\lambda_s^*)\over |\lambda_d|^2}\,
{{\rm Im}[(T^*+E^*+P_d^*+P\!E_d^*+P\!A_d^*)(P_s+P\!E_s+P\!A_s)]_{\pi\pi}\over |T+E|^2_{\pi\pi} }   \non \\
&\approx& 1.2\times 10^{-3} \left| {P_s+P\!E_s+P\!A_s\over T+E}\right|_{\pi\pi}\sin\delta_{\pi\pi},
\en
is a special case of Eq.~(\ref{eq:pipiacp}) with negligible $\Delta P$ and similarly for $a_{dir}^{({\rm t+p})}(K^+K^-)$.


Using the input parameters for the light-cone distribution amplitudes of light mesons, quark masses and decay constants from Refs.~\cite{CCBud,Laiho} and form factors from Refs.~\cite{ChengChiang,YLWu},
we find to the leading order expansion in $\Lambda_{\rm QCD}/m_b$ in QCDF that
\be \label{eq:Pover T}
&& \left({P_d\over T}\right)_{\pi\pi}=0.239\, e^{-i152^\circ}, \qquad \left({P_s\over T}\right)_{\pi\pi}=0.244\, e^{-i154^\circ}, \qquad
\left({\Delta P\over T}\right)_{\pi\pi}=0.010\, e^{-i35^\circ}, \non \\
&& \left({P_d\over T}\right)_{_{K\!K}}=0.235\, e^{-i152^\circ}, \qquad \left({P_s\over T}\right)_{_{K\!K}}=0.240\, e^{-i154^\circ}, \qquad
\left({\Delta P\over T}\right)_{_{K\!K}}=0.009\, e^{-i35^\circ}.
\en
Therefore, $\Delta P=P_d-P_s$ arising from the difference in the $d$- and $s$-loop penguin contractions [see Eq. (\ref{eq:penguinWC})] is very small compared to the tree amplitude. More precisely, it comes from the differences between $G_{\pi,K}(s_d)$ and $G_{\pi,K}(s_s)$ and between  $\hat G_{\pi,K}(s_d)$ and $\hat G_{\pi,K}(s_s)$ defined in Eq.~(\ref{eq:G}). Because of the smallness of $\Delta P$, we need to rely on SU(3) violation in both $T$ and $E$ amplitudes to explain the large disparity in the rates of $D^0\to K^+K^-$ and $\pi^+\pi^-$. It is straightforward to find
\be
\left({P_s\over T+E}\right)_{\pi\pi}=0.35\, e^{i170^\circ}, \qquad
\left({P_d\over T+E}\right)_{_{K\!K}}=0.24\, e^{i170^\circ}.
\en
Hence, $\delta_{\pi\pi}\approx \delta_{K\!K}=170^\circ$. From Eqs.~(\ref{eq:pipiacp}) and (\ref{eq:KKacp}), we derive
$a_{dir}^{\rm (t+p)}(\pi^+\pi^-)=6.7\times 10^{-5}$ and $a_{dir}^{\rm (t+p)}(K^+K^-)=-4.9\times 10^{-5}$. Therefore,   QCD-penguin induced \CP asymmetries in $D^0\to \pi^+\pi^-,~ K^+K^-$ are small mainly due to the almost trivial strong phases $\delta_{\pi\pi}$ and $\delta_{K\!K}$.

For QCD penguin power corrections, we shall consider weak penguin annihilation, namely, QCD-penguin exchange $\PE$ and QCD-penguin annihilation $\PA$.
At the short-distance level, weak penguin annihilation contributions are found to be smaller than QCD penguin with the hierarchy $P>\PE>\PA$. For example, $(\PE/T)_{\pi\pi}\sim 0.04$ and $(\PA/T)_{\pi\pi}\sim -0.02$.
In the QCDF approach, it can be shown that the short-distance weak annihilation and weak penguin annihilation terms are related to each other via (see Eq.~(55) of \cite{BN})
\be \label{eq:ann relation}
&& A^{\rm ^{SD}}={c_2\over c_1}E^{\rm ^{SD}}, \qquad  \PA^{\rm ^{SD}}={c_4+c_6\over c_2}A^{\rm ^{SD}}, \non \\
&& \PE^{\rm ^{SD}}={c_3\over c_1}E^{\rm ^{SD}}+{G_F\over\sqrt{2}}(if_Df_{M_1}f_{M_2})\left[c_5A_3^i+(c_5+N_c c_6)A_3^f\right],
\en
where $A_3^{i,f}$ denote annihilation amplitudes induced from $(-2)(S-P)\otimes (S+P)$ 4-quark operator and the superscripts `$i$' and `$f$' refer to gluon emission from the initial and final-state quarks, respectively. For example,
the amplitude $A_3$ has the expression $A_3\propto -2\la M_1M_2|(\bar uq)_{_{S+P}}\otimes(\bar qc)_{_{S-P}}|D\ra$ with $(\bar q_1q_2)_{_{S\pm P}}\equiv \bar q_1(1\pm\gamma_5)q_2$.
Now $A_3^f$ corresponds to the factorizable contribution
\be
A_3^f\propto -2\la M_1M_2|(\bar uq)_{_{S+P}}|0\ra\la 0|(\bar qc)_{_{S-P}}|D\ra \ ,
\en
where $A_3^i$ to the nonfactorizable contribution of $A_3$. The factorizable term $A_3^f$ was evaluated in \cite{Li2012} by assuming its dominance by lowest scalar resonances. In \cite{Kagan}, the large-$N_c$ counting
\be
{ \la M_1M_2|(\bar uu)_{_{S+P}}\otimes (\bar uc)_{_{S-P}}|D\ra \over \la M_1M_2|(\bar s_\alpha s_\beta-\bar d_\alpha d_\beta)_{_{V-A}}\otimes (\bar u_\beta c_\alpha)_{_{V-A}}|D\ra}={\cal O}(N_c)
\en
was employed to get the relation $\PE \sim (2N_c c_6^{\rm eff}/c_1)E$. (The amplitude $\PE$ was denoted by $P_{f,1}$ in \cite{Kagan}.)
However, the major contributions to weak annihilation and weak penguin annihilation from FSI's were not considered in \cite{Kagan,Li2012}. \footnote{Contributions from final-state interactions have the general expression
$\sum_n \la M_1M_2|{\cal L}_S|n\ra\la n|{\cal L}_W|D\ra$, where ${\cal L}_S$ and ${\cal L}_W$ are strong- and weak-interaction Lagrangians, respectively, and $n$ denotes the physical intermediate states.
}
We would like to stress again that the relations given in Eq.~(\ref{eq:ann relation}) are valid only for short-distance ones. One cannot replace $E^{\rm ^{SD}}$ and $A^{\rm ^{SD}}$ by the topological amplitudes $E$ and $A$, respectively, extracted from the data.
Moreover, annihilation terms at short-distance level are small due to helicity suppression. Typically, QCDF yields $E^{\rm ^{SD}}\sim 0.5$ and $A^{\rm ^{SD}}\sim -0.2$ in units of $10^{-6}$ GeV. Comparing with the topological amplitudes extracted in Eq.~(\ref{eq:PP1}), it is evident that weak annihilation amplitudes are dominated by long-distance contributions.

As pointed out in \cite{Cheng:2012},
long-distance contributions to SCS decays, for example, $D^0\to \pi^+\pi^-$, can proceed through the weak decay $D^0\to K^+K^-$ followed by a resonant-like final-state rescattering as depicted in Fig.~2 of \cite{Cheng:2012}. It has the same topology as the QCD-penguin exchange topological graph $P\!E$.  Just as the weak annihilation topologies $E$ and $A$, it is expected that weak penguin annihilation will receive sizable long-distance contributions from FSI's as well. Recall that soft corrections due to penguin annihilation have been proposed to resolve some problems in hadronic $B$ decays, such as the rate deficit problem for penguin-dominated decays and the \CP puzzle for $\bar B^0\to K^-\pi^+$ \cite{BN}. Hence, it is plausible to assume that $P\!E$ is of the same order of magnitude as $E$. For concreteness, we shall take (in units of $10^{-6}$ GeV)
\be \label{eq:PE}
P\!E=1.6\,e^{i 115^\circ} \ .
\en

As shown in Table~\ref{tab:CPVpp}, we see that the predicted \CP violation denoted by $a_{dir}^{(\rm tot)}$ or $a_{dir}^{(\rm tree)}$
is at most of order $10^{-3}$ in the SM.  For $\Delta a_{CP}^{\rm dir}$, the \CP asymmetry difference between $D^0\to K^+K^-$ and $D^0\to \pi^+\pi^-$, we obtain $-(0.139\pm0.004)\%$ and $-(0.151\pm0.004)\%$ for Solutions I and II, respectively. It is of interest to notice that the prediction of $-0.118\%$ obtained in \cite{Li2012} based on a different approach is in agreement with our result.
Since in the SM, $\Delta a_{CP}^{\rm dir}$ arises mainly from weak penguin annihilation, we can vary the amplitude $P\!E$ to see how much enhancement we can gain. Even with the maximal magnitude $|P\!E|\sim T$ and a maximal strong phase relative to $T$, we get $\Delta a_{CP}^{\rm dir}=-0.27\%$. This is more than $2\sigma$ away from the current world average. Hence, if the LHCb result for $\Delta a_{CP}^{\rm dir}$ is confirmed by further data analysis, it will imply new physics in the charm sector.

\subsection{$D^+\to\pi^+\pi^0$}

It is known that the $D^+\to\pi^+\pi^0$ decay does not receive QCD penguin contributions and its direct  \CP violation vanishes. Nevertheless, it does receive additional contributions from isospin violation and electroweak penguin effects. For example, the $u$-$d$ quark mass difference will induce $\pi$-$\eta$-$\eta'$ mixing. Hence, the decay $D^+\to\pi^+\eta$ ($D^+\to\pi^+\eta'$) followed by the $\eta$-$\pi$ ($\eta'$-$\pi$) mixing will contribute to the direct \CP asymmetry in $D^+\to\pi^+\pi^0$. Consider the contribution from $\pi^0$-$\eta_8$ mixing
\be
A(D^+\to \pi^+\pi^0)={1\over\sqrt{2}}\lambda_d(T+C)_{\pi\pi}+\la \pi^0|{\cal H}_{\rm mass}|\eta_8\ra{1\over m_\pi^2-m_\eta^2}A(D^+\to\pi^+\eta) \ ,
\en
where (see {\it e.g.}, \cite{Donoghue})
\be
\la \pi^0|{\cal H}_{\rm mass}|\eta_8\ra=-{\sqrt{3}\over 4} {m_d-m_u\over m_s-\hat m}(m_\eta^2-m_\pi^2)
\en
with $\hat m=(m_u+m_d)/2$. It follows that
\be
A(D^+\to \pi^+\pi^0)={1\over\sqrt{2}}\lambda_d(T+C)_{\pi\pi}+{\sqrt{3}\over 4}\,{m_d-m_u\over m_s-\hat m}A(D^+\to\pi^+\eta) \ .
\en
There will be also a correction from the $\eta'$ meson.
Consequently, isospin breaking induced by $u$ and $d$ quark mass difference will generate \CP asymmetries at tree and loop levels for $D^+\to \pi^+\pi^0$. However, it is suppressed by a factor of $(m_d-m_u)/m_s\approx 0.025$.  Therefore, \CP asymmetry in $D^+\to\pi^+\pi^0$ induced by isospin breaking and electroweak penguins are negligible. Recently, it was argued in \cite{Zupan} that a measurement of nonzero $a_{CP}^{\rm dir}(\pi^+\pi^0)$ would be a signal for $\Delta I=3/2$ new physics.

\section{New Physics Effects \label{sec:NP}}

Whether the first evidence of \CP violation in the charm sector observed by LHCb is consistent with the SM or implies NP will require a further analysis of more data and improved theoretical understanding.
If the improved theoretical estimate of $\Delta a_{CP}^{\rm dir}$ remains to be a few per mille and the experimental measurement continues to be large with more statistics in the future or if the direct \CP asymmetry of any of the discussed modes is significantly larger than $10^{-3}$, it will be clear evidence of physics beyond the SM in the charm sector. Then it will be important to explore possible NP scenarios responsible for such large direct \CP asymmetries.

\subsection{Large penguins}

In the wake of the LHCb and CDF measurements of $\Delta a_{CP}^{\rm dir}$, several groups of people have assumed enhanced hadronic matrix elements through SU(3) breaking \cite{PU} or $U$-spin breaking \cite{Feldmann} or enhanced penguins via unforeseen QCD or NP effects \cite{Brod,Bhattacharya:2012,Altmannshofer}. Let us examine the phenomenological implications of large penguins irrespective of the origin of enhancement. It turns out that the penguin
\be \label{eq:largeP}
{1\over 2}\Sigma P\equiv {1\over 2}(P_d+\PE_d+\PA_d+P_s+\PE_s+\PA_s)\approx
\left\{
\begin{array}{ll}
2.9\,T e^{i 85^\circ} & \mbox{for Solution I} ~, \\
3.2\,T e^{i 85^\circ} & \mbox{for Solution II} ~,
\end{array}
\right.
\en
with maximal strong phase can accommodate the measurement of direct \CP asymmetry difference between $D^0\to K^+K^-$ and $\pi^+\pi^-$ (see Table~\ref{tab:CPVpp_new}). Indeed, it has been noticed in \cite{Bhattacharya:2012,Brod} that a large penguin of order $3T$ can explain the observed $\Delta a_{CP}^{\rm dir}$. \footnote{Authors of \cite{Bhattacharya:2012} have introduced an additional phenomenological penguin amplitude $P_b$ in order to accommodate the measured $\Delta a_{CP}^{\rm dir}$.  However, there is no need to make this assumption as the penguin amplitudes can be recast to
$\lambda_d P_d+\lambda_s P_s={1\over 2}(\lambda_d-\lambda_s)(P_d-P_s)-{1\over 2}\lambda_b(P_d+P_s)$.  The last term on the right-hand side of the above relation is the so-called $P_b$ in \cite{Bhattacharya:2012}.}
Some authors argued that penguin matrix elements (or more pertinently ``penguin amplitudes") could be substantially enhanced in the SM just like the enhancement of the $W$-exchange amplitude relative to the naive expectation. However, this conjecture needs to be clarified. The topological amplitudes $C$ and $E$ extracted from the data are much larger than what expected from naive factorization. This is because they receive large $1/m_c$ power corrections from FSI's. For example, the topological amplitude $E$ receives contributions from the tree amplitude $T$ via final-state rescattering. As shown in Refs.~\cite{Zen,Chenga1a2}, the effects of final-state rescattering via nearby resonances can be quantified. As for the QCD penguin, an estimate to the leading-order expansion of $\Lambda_{\rm QCD}/m_b$ in QCDF is given in Eq.~(\ref{eq:Pover T}). The $1/m_c$ corrections manifested as weak penguin annihilation are even smaller. Nevertheless, weak penguin annihilation does receive long-distance contributions from the color-allowed tree amplitude followed by final-state rescattering. Since this FSI originates from the tree amplitude, one cannot have $\PE>T$. Hence, we believe that a penguin amplitude larger than the tree amplitude in size is only possible through the enhancement of new physics.


Using the above large penguin ${1\over 2}\Sigma P$ as input, the predicted direct \CP asymmetries for other charm decays are summarized in the second column of Table~\ref{tab:CPVpp_new}. We see that many modes, such as $D_s^+\to \pi^+K^0,\pi^0 K^+,K^+\eta'$, are expected to yield direct \CP asymmetries of a similar magnitude, at a few per mille level.


\begin{table}
\caption{Direct \CP asymmetries (in units of $10^{-3}$) of SCS $D\to PP$ decays estimated in the scenarios with large penguin contributions  and large chromomagnetic dipole operator. The parameters $\Sigma P$ and $c_{8g}^{\rm NP}$ are chosen to fit the data of $\Delta a_{CP}^{\rm dir}$:
${1\over 2}\Sigma P=2.9\,Te^{i85^\circ}$ and $c_{8g}^{\rm NP}=0.017e^{i14^\circ}$ for Solution I,  ${1\over 2}\Sigma P=3.2\,Te^{i85^\circ}$ and  $c_{8g}^{\rm NP}=0.012e^{i14^\circ}$ for Solution II. The number in parentheses
is for Solution II of $E_d$ and $E_s$ [Eq.~(\ref{eq:EdEs})].
  \label{tab:CPVpp_new}}
\begin{tabular}{ l   c c} \hline\hline
Decay Mode~~ & Large penguins~~~ & Large c.d.o. \\
 \hline
$D^0\to \pi^+ \pi^-$ & 3.96 (4.40)~~~ & 5.18 (3.70) \\
$D^0\to\pi^0 \pi^0$   & 0.93 (1.01)~~~ & 8.63 (6.19)  \\
$D^0\to \pi^0 \eta $  & 0.09 (0.03)~~~ & $-6.12$ ($-4.15$) \\
$D^0\to \pi^0 \eta'$  & 2.36 (2.67)~~~ & $-0.44$ ($-0.44$) \\
$D^0\to \eta\eta $ &  $-1.79$ ($-1.64$)~~~ & $-1.63$ ($-2.00$)\\
$D^0\to \eta\eta' $  & $2.65$ (1.49)~~~  & $-2.30$ ($-1.08$) \\
$D^0\to K^+ K^{-}$   & $-2.63$ ($-2.36$)~~~ & $-1.46$ $(-2.88)$ \\
$D^+\to \pi^+ \pi^0$   & 0 (0)~~~ & 0 (0) \\
$D^+\to \pi^+ \eta $  &  $-3.24$ $(-3.62)$~~~ & $-5.35$ $(-3.67)$   \\
$D^+\to\pi^+ \eta' $  & $2.97$ (3.34)~~~ &  0.93 (0.59) \\
$D^+\to K^+ \ol{K}^{0}$  &  $-2.95$ ($-3.28$)~~~ & 0.37 (0.29)  \\
$D_s^+\to \pi^+ K^{0}$  & $3.29$ (3.66)~~~ & $-0.47$ $(-0.35)$  \\
$D_s^+\to \pi^0 K^{+}$  &  $4.57$ (5.08)~~~ & 4.40 (3.14)\\
$D_s^+\to K^{+}\eta$   &  $-0.58$ ($-0.57$)~~~ & 1.59 (0.94) \\
$D_s^+\to K^{+}\eta' $  &  $-5.16$ ($-5.79$)~~~ & 1.76 (1.39)\\
\hline\hline
\end{tabular}
\end{table}
%

\subsection{Large chromomagnetic dipole operator}

Even before the LHCb experiment, the impact of NP on $\Delta a_{CP}^{\rm dir}$ had already been investigated in \cite{Grossman,BigiCDF}. The unexpected LHCb measurement has inspired many analyses based on a variety of NP models. Models for NP effects at tree level include flavor-changing coupling of a SM $Z$ boson \cite{Giudice,Altmannshofer}, flavor-changing neutral currents induced by a leptophobic massive $Z'$ boson \cite{Zhu,Altmannshofer}, two Higgs-doublet model \cite{Altmannshofer}, color-singlet scalar model \cite{Nir}, color-sextet scalar ({\it i.e.}, diquark scalar) model \cite{Altmannshofer,Chen}, color-octet scalar model \cite{Altmannshofer} and fourth generation model \cite{Rozanov,Feldmann}. Models with NP in QCD penguins at the loop level have been constructed as well, including new fermion and scalar fields \cite{Altmannshofer} and the chirally enhanced chromomagnetic dipole operator \cite{Giudice}.

The NP models are highly constrained by $D^0$-$\ov D^0$ mixing, $K^0$-$\ov K^0$ mixing and \CP violation in the kaon system characterized by the parameter $\epsilon'/\epsilon$ \cite{Ligeti}. Many of the tree-level NP models are either ruled out or in tension with experiments \cite{Altmannshofer}. As stressed in \cite{Giudice}, a large NP contribution to the $\Delta C=1$ chromomagnetic dipole operator is the best candidate to explain the LHCb and CDF results as it is least constrained by all current data in flavor physics. Although the chromomagnetic dipole operator $O_{8g}$ is suppressed by the charm Yukawa coupling, the hadronic matrix element reads
\be
\la M_1M_2|O_{8g}|D\ra=-{\alpha_s\over \pi}\,{m_c\over k^2}\la M_1M_2|\bar u_\alpha\gamma_\mu /\!\!\!k(1+\gamma_5){\lambda_{\alpha\beta}\over 2}c_\beta\,\bar q_\gamma\gamma^\mu{\lambda_{\gamma\delta}\over 2}q_\delta|D\ra
\en
where $k^2$ is the square of momentum transfer of the gluon and is of order $m_c^2$. After applying the equation of motion, we see that the matrix element $\la M_1M_2|O_{8g}|D\ra$ is independent of $m_c$; that is, it is enhanced by a factor of $v/m_c$, where $v$ is the vacuum expectation value of the Higgs field arising from the structure of the gauge-invariant dimension-six operator \cite{Giudice}.
On the other hand, the $D^0$-$\ov D^0$ mixing induced by $O_{8g}$ is suppressed by a factor of $m_c^2/v^2$. Of course, we need NP to enhance the Wilson coefficient $c_{8g}$ and to induce a sizable imaginary part. This can be realized in the supersymmetric models where the gluino-squark loop contributes a major part of $c_{8g}$ \cite{Kagan}, the disoriented $A$ terms and split families are the sources of flavor violation \cite{Giudice}, or the flavor structure of the trilinear scalar couplings is related to the structure of the Yukawa couplings via approximate flavor symmetries \cite{Hiller}.

To demonstrate the NP effects, we consider the NP penguin amplitude $P^{\rm NP}$ induced by the chromomagnetic operator which has the expression
\be
P^{\rm NP}_{P_1P_2} &=& {G_F \over \sqrt{2}} [a^{\rm NP}_4(P_1P_2)+r_\chi^{P_2} a^{\rm NP}_6(P_1P_2)]X^{(DP_1,P_2)} \ ,
\en
with
\be
a_4^{\rm NP}(P_1P_2) &=& {\cal P}^{\rm NP}_4(P_2)=-2c_{8g}^{\rm NP}{C_F\alpha_s\over 4\pi N_c}\int^1_0 {dx\over 1-x}\Phi_{P_2}(x),  \non \\
a_6^{\rm NP}(P_1P_2) &=& {\cal P}^{\rm NP}_6(P_2)=-2c_{8g}^{\rm NP}{C_F\alpha_s\over 4\pi N_c}.
\en
See Eqs.~(\ref{eq:penguinWC}) and (\ref{eq:P46}) for a derivation.
Then we add the NP amplitude $P^{\rm NP}$  to the penguin amplitudes $\lambda_p P_p$ in Table~\ref{tab:PPSCSamp} where summation over $p=d,s$ is understood; that is, the penguin amplitudes $\lambda_p P_p$ are replaced by $\lambda_p P_p+P^{\rm NP}$.
As an example of illustration, we shall take $c_{8g}^{\rm NP}\approx 0.012e^{i14^\circ}$ which fits to the data of $\Delta a_{CP}^{\rm dir}$. The calculated \CP asymmetries for the other modes are listed in the last column of Table~\ref{tab:CPVpp_new}. Since the decay $D^0\to K^0\ov K^0$ does not receive QCD penguin contribution, it is not affected by the chromomagnetic dipole operator in NP.
It is interesting to notice that while a large chromomagnetic dipole operator leads to a large direct \CP asymmetry for $D^0\to\pi^0\pi^0,\pi^0\eta$, the predicted \CP violation for $D^0\to \pi^0\eta'$, $D^+\to\pi^+\eta', K^+\ov K^0$ and $D_s^+\to \pi^+K^0,K^+\eta'$ is much smaller than that in the large penguin scenario. Therefore, measurements of the \CP asymmetries of the above-mentioned modes will enable us to discriminate between the two different NP scenarios.

\section{Discussions and Conclusions \label{sec:summary}}

In this work we have examined various sources responsible for the seemingly large SU(3) breaking effects in the $D^0\to K^+K^-$ and $D^0\to\pi^+\pi^-$ decays. We considered three cases: (i) SU(3) symmetry holds for $T$ and $E$ amplitudes. Then a sizable  $\Delta P$ (the difference of $s$- and $d$-quark penguin contractions) of the same order of magnitude as $T$ is needed to explain the data. This is also true if the symmetry breaking of $T$ and $E$ amplitudes follows the pattern given by by Eq. (\ref{eq:UbreakinT}). (ii) SU(3) symmetry holds for $E$ amplitudes, but SU(3) violation due to decay constants, meson masses and form factors is taken into account in $T$ amplitudes so that $T_{_{K\!K}}/T_{\pi\pi}\approx 1.32$ as inferred from the factorization approach. This leads to $|\Delta P/T|\sim 0.15$\,. (iii) The large rate difference between $D^0\to K^+K^-$ and $D^0\to\pi^+\pi^-$ is entirely accounted for by SU(3) violation in $T$ and $E$ amplitudes and hence $\Delta P$ can be neglected. Owing to the observation of $D^0\to K^0\ov K^0$ through $W$-exchange and penguin annihilation diagrams and the smallness of $\Delta P$ theoretically, we have argued in this work that the last scenario is preferred and fixed the SU(3) breaking in the $W$-exchange amplitudes from the following four modes: $K^+K^-$, $\pi^+\pi^-$, $\pi^0\pi^0$ and $K^0\ov K^0$. Our results are summarized in Table~\ref{tab:BFpp}, where we have shown branching fractions of SCS $D\to PP$ decays and elaborated on SU(3) breaking effects for each mode.


Since the magnitude and the phase of topological color-suppressed tree amplitude and weak annihilation amplitudes which arise mainly from final-state rescattering can be extracted from the data in the diagrammatic approach, direct \CP asymmetry $a_{dir}^{\rm (tree)}$ at tree level can be reliably estimated. We predict that $a_{dir}^{\rm (tree)}(D^0\to K^0\ov K^0)$ ranges from $-0.73\times 10^{-3}$ to $-1.73\times 10^{-3}$, depending on the solution for $E$ amplitudes. A recent similar study in \cite{Li2012} found opposite signs of $a_{dir}^{\rm (tree)}$ to ours for most of the SCS $D\to PP$ decays. This is ascribed to the phase of $E$ amplitudes: It is in the second quadrant in our work while in the third quadrant in \cite{Li2012}. As for the decay $D^+\to \pi^+\pi^0$, it does receive corrections from isospin violation due to the  $u$-$d$ quark mass difference and electroweak penguins. However, the induced \CP asymmetry is negligible because, for example, the isospin-violating effect is suppressed by a factor of $(m_d-m_u)/m_s$.

Using QCD factorization as a guideline, the direct \CP asymmetries of both $D^0\to K^+K^-$ and $D^0\to\pi^+\pi^-$ are at a few $\times 10^{-4}$ level.  This is seen to be largely due to the trivial relative strong phase between the QCD penguin amplitude and the tree-level amplitudes. For QCD penguin power corrections, the short-distance contributions to weak penguin annihilation diagrams $\PE$ and $\PA$ are small, but $\PE$ receives long-distance final-state contributions from rescattering through nearby resonances which have the same topology as the $W$-exchange diagram. It is thus natural to assume that $\PE$ is of the same order of magnitude as $E$. We conclude that the \CP asymmetry difference $\Delta a_{CP}^{\rm dir}$ between $D^0 \to K^+ K^-$ and $D^0 \to \pi^+ \pi^-$ is about $-(0.139\pm 0.004)\%$ and $-(0.151\pm 0.004)\%$ for the two solutions of $E$ amplitudes, respectively. A similar prediction was also obtained in \cite{Li2012}.

If $\Delta a_{CP}^{\rm dir}$ continues to be large with more statistics in the future, it will be clear evidence of physics beyond the standard model in the charm sector. Considering two possibilities of new physics effects, namely, large penguins and large chromomagnetic dipole operator, we have studied their phenomenological implications in the SCS charmed meson decays. We point out that the \CP asymmetries several modes, such as $D^0\to\pi^0\pi^0,\pi^0\eta^{(')}$, $D^+\to\pi^+\eta', K^+\ov K^0$ and $D_s^+\to \pi^+K^0,K^+\eta'$, allow us to discriminate between different new physics scenarios.

\section*{Acknowledgments}

We are grateful to Alexander Kagan, Hsiang-nan Li and Fu-Sheng Yu for useful discussions.  This research was supported in part by the National Science Council of Taiwan, R.~O.~C. under Grant Nos.~NSC-100-2112-M-001-009-MY3 and NSC-100-2628-M-008-003-MY4.


\end{document}